\documentclass[twocolumn,aps,prc,superscriptaddress,showpacs,nofootinbib,floatfix]{revtex4}
\usepackage{epsfig,bm,feynmf}
\usepackage{graphicx}
\usepackage{amsmath}
\usepackage{dcolumn}
\usepackage{graphicx}
\usepackage[usenames,dvipsnames,svgnames,table]{xcolor}

\usepackage{tikz-feynman,contour}
\tikzfeynmanset{compat=1.0.0}
\usepackage[normalem]{ulem}  

\renewcommand\sout{\bgroup \color{red} \ULdepth=-.5ex \ULset}

\begin{document}

\title{Heavy quark coalescence probability in the presence of a potential}

\author{Taesoo Song}\email{t.song@gsi.de}
\affiliation{GSI Helmholtzzentrum f\"{u}r Schwerionenforschung GmbH, Planckstrasse 1, 64291 Darmstadt, Germany}

\author{Jiaxing Zhao}\email{jzhao@itp.uni-frankfurt.de}
\affiliation{Institute for Theoretical Physics, Johann Wolfgang Goethe Universit\"{a}t, Frankfurt am Main, Germany}
\affiliation{Helmholtz Research Academy Hessen for FAIR (HFHF),GSI Helmholtz Center for Heavy Ion Research. Campus Frankfurt, 60438 Frankfurt, Germany}

\begin{abstract}
In this study, we explore the role of the heavy-quark potential in heavy-quark coalescence, whose probability is expected to be unity at low momentum.
To this end, we develop a phenomenological heavy-quark potential based on the constituent quark model that reproduces the vacuum masses of pseudoscalar and vector heavy mesons. Using this potential, we demonstrate its enhancement effect on the coalescence probability.
We also investigate how medium-induced modifications of the heavy-quark potential in the quark–gluon plasma (QGP) affect the coalescence process. Our results indicate that the coalescence probability remains close to unity as long as the modification of the potential is sufficiently moderate.
\end{abstract}

\maketitle

\section{Introduction}
Relativistic heavy-ion collisions create a hot and dense state of QCD matter known as the quark-gluon plasma (QGP).
The properties of the QGP can be investigated using heavy-flavor which are predominantly produced in the initial hard scatterings between nucleons~\cite{Rapp:2009my,Andronic:2015wma,Dong:2019unq,Dong:2019byy,Zhao:2020jqu}.
At sufficiently high collision energies, such as those of the LHC, thermal production of charm quarks may become possible.
However, current experimental measurements at the LHC provide no evidence for thermal charm production and instead support a picture in which charm quarks are massive throughout the QGP phase~\cite{Song:2024hvv}.

Heavy flavors have several advantages as probes of the QGP.
Because they are produced in hard scattering processes, heavy quarks are generated on very short time scales and therefore experience the entire evolution of the medium.
Owing to their large masses, they retain information about the early stages of the collision, which is not completely erased by subsequent interactions with the hot and dense medium.
Furthermore, their production can be calculated within perturbative QCD (pQCD), providing a relatively model-independent theoretical framework.

After their production, heavy quarks interact with medium constituents through both elastic and inelastic (radiative) processes. 
Elastic scatterings dominate at relatively low energies, whereas radiative energy loss becomes increasingly important at higher energies and eventually overtakes the elastic contribution~\cite{Grishmanovskii:2025mnc}.

As the system expands and cools, the QGP undergoes a transition from a partonic to a hadronic phase through hadronization.
Since hadronization is intrinsically non-perturbative, its description relies on phenomenological models. 
One of the most successful approaches is the parton coalescence (recombination) model~\cite{Fries:2003kq,Greco:2003mm,Greco:2003xt,Plumari:2017ntm,Cao:2019iqs,Greco:2003vf,He:2019vgs,Song:2015sfa,Song:2015ykw,Zhao:2024ecc}, which naturally explains the constituent-quark-number scaling of elliptic flow and the larger elliptic flow coefficient $v_2$ observed for baryons relative to mesons at intermediate transverse momentum $p_T$~\cite{STAR:2003wqp,PHENIX:2003qra}.
At high $p_T$, however, heavy-quark hadronization is expected to be dominated by fragmentation, in which a heavy quark forms a hadron through the emission of soft gluons.

Fragmentation and coalescence affect heavy-quark kinematics in opposite ways.
During fragmentation, a heavy quark loses energy and momentum, whereas in coalescence it gains energy and momentum from its recombination partners.
Consequently, coalescence enhances collective flow effects in heavy-ion collisions.
In many phenomenological studies, the coalescence probability of a heavy quark is small at high $p_T$ and increases as $p_T$ decreases, see review papers~\cite{Rapp:2018qla,Zhao:2023nrz,Altmann:2024kwx}.
In the limit $p_T\rightarrow 0$, the coalescence probability is generally assumed to approach unity, since a static heavy quark cannot hadronize through fragmentation.

In principle, the coalescence probability for each heavy meson and heavy baryon state depends on its own characteristic radius.
Because of the large number of hadronic states involved, determining suitable radii for all states is impractical.
Therefore, one can focus on the ground state, $D$ or $B$ meson and assume that the coalescence probabilities for excited states follow the hadron yield ratios predicted by statistical hadronization model~\cite{Song:2021mvc}.
This procedure allows one to estimate the total coalescence probability of a heavy quark.

However, it has been shown that the total heavy-quark coalescence probability remains significantly below unity within the quasi-particle model framework~\cite{Song:2021mvc}.
The reason is that the density of light (anti)quarks at the critical temperature $T_c$, constrained to reproduce the lattice-QCD equation of state, is too low to yield a sufficient large coalescence probability, even when the light (anti)quarks are treated as massless. 
Consequently, phenomenological models often introduce an ad hoc normalization factor to rescale the total coalescence probability such that it reaches unity at $p_T=0$~\cite{Song:2021mvc}.

Recently we demonstrated that incorporating the heavy-quark potential in the coalescence model for quarkonium production yields results consistent with those obtained from the statistical model~\cite{Song:2025zfy}.
In the weak-binding limit, the coalescence and statistical models generally predict similar particle yields.
For strongly bound states, however, the bound state mass is significantly smaller than the sum of the constituent masses, causing the conventional coalescence model to predict lower yields than the statistical model.
When the interaction potential is included, the spatial distribution of constituent (anti)quarks becomes nonuniform and is enhanced in the vicinity of potential coalescence partners.
This increases the coalescence probability increases and restores consistency between the coalescence and statistical approaches.

In the present study, we apply the same idea to address the problem of the insufficient heavy-quark coalescence probability near $T_c$.
To this end, we introduce phenomenological heavy-quark potentials that reproduce the vacuum masses of heavy mesons.
Medium modifications of the interaction are incorporated through a temperature-dependent screening mass.

This paper is organized as follows.
In Sec.~\ref{potentialQq}, we construct heavy-light quark potentials that reproduce the observed heavy-meson masses.
The resulting total coalescence probability is calculated in Sec.~\ref{coal-pro}.
In Sec.~\ref{in-medium}, we investigate medium modifications of the potential and their impact on the coalescence probability.
Finally, a summary is presented in Sec.~\ref{summary}.

\section{Heavy meson potential}
\label{potentialQq}
For the interaction between a heavy quark and a light antiquark forming a heavy meson, we adopt a heavy-light quark potential motivated by lattice-QCD calculations of the heavy-quark free energy~\cite{Satz:2005hx,Digal:2005ht,Kaczmarek:2008saj},
\begin{eqnarray}
F(r,T)=F_c(r,T)+F_s(r,T),
\label{freeE}
\end{eqnarray}
where $F_c$ and $F_s$ denote the Coulombic and string contributions, respectively, and are given by
\begin{eqnarray}
F_c(r,T)&=&-\frac{\alpha_s}{r}[e^{-\mu r}+\mu r],\\
F_s(r,T)&=&\frac{\sigma}{\mu}\bigg[\frac{\Gamma(1/4)}{2^{3/2}\Gamma(3/4)}-\frac{\sqrt{\mu r}}{2^{3/4}\Gamma(3/4)}K_{1/4}[(\mu r)^2]\bigg]\nonumber,
\label{FcFs}
\end{eqnarray}
with $\alpha_s$, $\sigma$ and $\mu$ representing the effective strong coupling, string tension, and screening mass, respectively~\cite{Satz:2005hx}.

This potential has several attractive features.
First, it is motivated by lattice-QCD calculations.
Second, medium effects can be incorporated naturally through the temperature dependence of the screening mass $\mu$.

To improve the description of the heavy-meson spectrum, we include a spin-spin interaction term~\cite{Bhaduri:1981pn,Park:2016xrw}.
The resulting heavy-light quark potential is
\begin{eqnarray}
V(r,T)=F(r,T)+\alpha_s\frac{\sigma_Q\cdot \sigma_q}{m_Q m_q}\frac{e^{-r/r_0}}{r_0^2 r},
\label{potential}
\end{eqnarray}
where the factor $e^{-r/r_0}/(r_0^2r)$ represents a finite-size form factor and reduces to $4\pi\delta(r)$ in the limit $r_0 \rightarrow 0$~\cite{Bhaduri:1981pn}.
The spin factor is given by
\begin{eqnarray}
\sigma_Q\cdot \sigma_q=\frac{S(S+1)-3/2}{2},
\end{eqnarray}
where $S$ is the total spin of the heavy meson.
The parameter $r_0$ is defined as  
\begin{eqnarray}
r_0=\bigg(\alpha+\beta\frac{m_Q m_q}{m_Q+m_q}\bigg)^{-1}
\end{eqnarray}
with $\alpha=2.2~ fm^{-1}$ and $\beta=0.277$~\cite{Park:2016xrw}.

Throughout this work, the constituent quark mass is fixed at $m_q=0.33 $ GeV, while the heavy-quark mass $m_Q$ and the coupling constant $\alpha_s$ are treated as free parameters and determined by fitting the masses of the $D(B)$ and $D^*(B^*)$ mesons.

The potential is then employed in the Schr\"{o}dinger
equation for a heavy quark - light anti-quark system,
\begin{eqnarray}
\bigg[-\frac{\nabla^2}{2m_\mu}+\widetilde{V}(r,T)\bigg]\psi(r,T)
=-\varepsilon\psi(r,T),
\end{eqnarray}
where $m_\mu=m_Qm_q/(m_Q+m_q)$ is the reduced mass and $\widetilde{V}(r,T)\equiv V(r,T)-V(r=\infty, T)$ is the potential shifted to vanish at infinite separation. 
Here $\psi(r,T)$ denotes the heavy meson wave function at temperature $T$ and $\varepsilon$ is the binding energy.
The heavy-meson mass is then given by 
\begin{eqnarray}
M = m_Q+m_q + V(r=\infty,T) - \varepsilon.
\end{eqnarray}

\begin{figure}[h]
\centerline{
\includegraphics[width=9 cm]{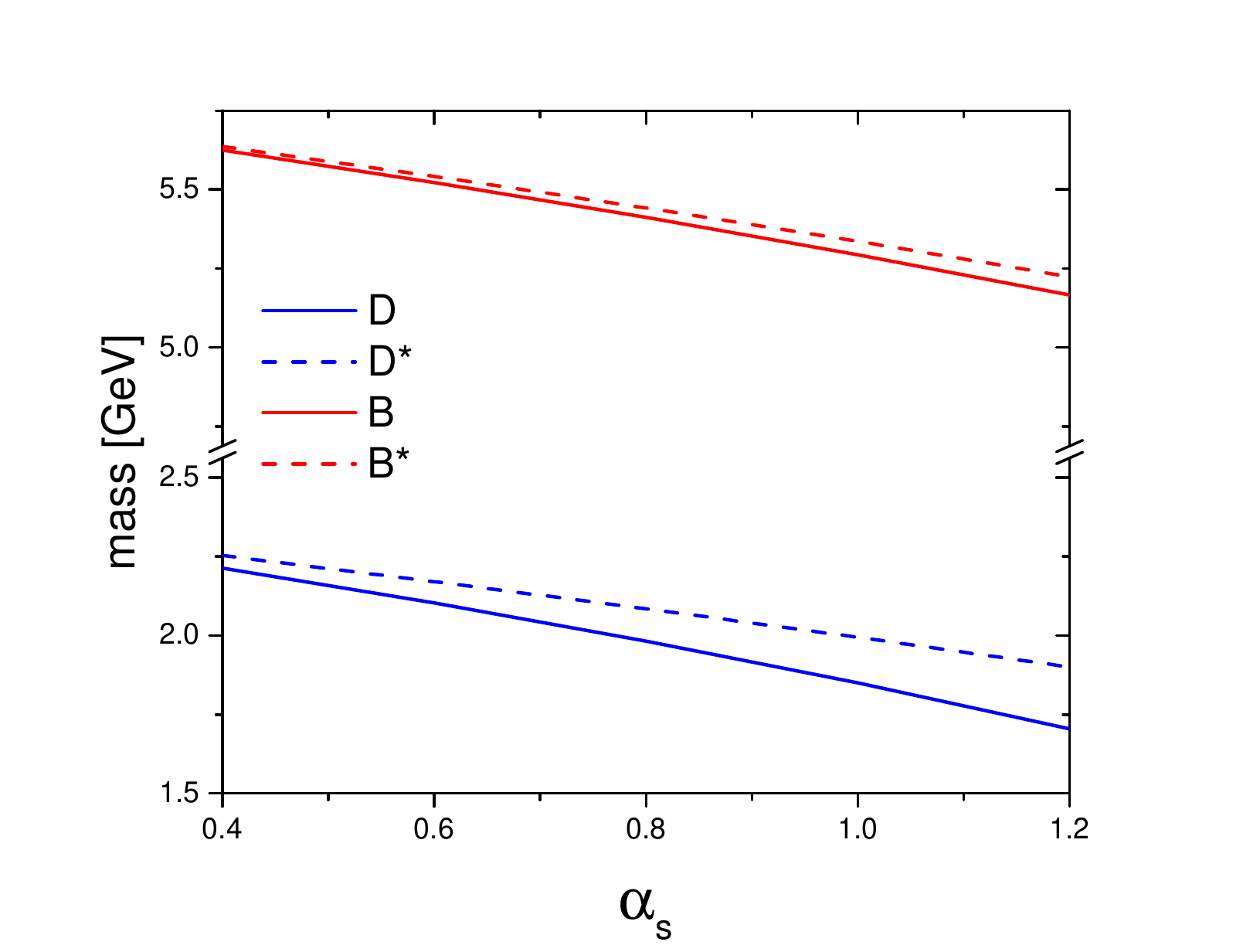}}
\caption{Masses of $D$, $D^*$, $B$ and $B^*$ mesons in vacuum ($T=0$) as functions of the strong coupling $\alpha_s$ for $m_c=$ 1.14 GeV and $m_b=$ 4.59 GeV.}
\label{alpha_s}
\end{figure}

Fig.~\ref{alpha_s} shows the vacuum masses of the  
heavy mesons in vacuum ($T=0$) as functions of the strong coupling $\alpha_s$ for $m_c=$ 1.14 GeV and $m_b=$ 4.59 GeV.
Since the string term in Eq.~(\ref{freeE}) is expected to be largely insensitive to the quark flavor, we adopt the same parameters relevant as in Ref.~\cite{Satz:2005hx,Gubler:2020hft}, namely $\sigma=0.445~{\rm GeV^2}$ and $\mu$=0.35$\sqrt{\sigma}$ in vacuum ($T=0$). 

The best agreement with the observed heavy-meson masses is obtained near $\alpha_s=$ 1 (0.97 for the charm sector and 1.03 for the bottom sector).
These values are significantly larger than the value $\alpha_s=\pi/12$ used for quarkonium~\cite{Satz:2005hx}.
This difference is expected because open heavy-flavor mesons are spatially more extended than quarkonia and therefore probe longer-distance interactions. 
We also note that $\alpha_s$ primarily controls the hyperfine mass splitting between the $D (B)$ and $D^* (B^*)$ states, whereas the spin-averaged meson masses are mainly determined by the heavy-quark mass $m_Q$.

\section{heavy quark coalescence probability}
\label{coal-pro}

Heavy quarks hadronize in heavy-ion collisions through either fragmentation or coalescence~\cite{Zhao:2023nrz}.
Fragmentation dominates at high $p_T$, whereas coalescence becomes the primary hadronization mechanism at low $p_T$.
Consequently, a static heavy quark is expected to hadronize exclusively through coalescence.

Within the coalescence model, the yield of hadrons produced through the process $1+2\rightarrow 3$ is given by~\cite{Song:2016lfv}
\begin{eqnarray}
N_3&=&\frac{1}{(2\pi)^6}\frac{D_3}{D_1D_2}\int d^3{\bf r_1}d^3{\bf r_2}d^3{\bf p_1}d^3{\bf p_2}\nonumber\\
&\times& f_1({\bf r_1,p_1})f_2({\bf r_2,p_2})\Phi(r,k),
\label{coal}
\end{eqnarray}
where $D_i$ denotes the color-spin degeneracy factor of particle $i$, and $\Phi(r,k)$ is the Wigner function of the produced hadron.

For a hadron in the 1-S state with a Gaussian wavefunction, the Wigner function takes the form   
\begin{eqnarray}
\Phi(r,k)=8\exp\bigg[-\frac{r^2}{\sigma^2}-\sigma^2k^2\bigg],
\label{wignerF}
\end{eqnarray}
where $r=|{\bf r_1-r_2}|$ and $k=|m_2{\bf p_1}-m_1{\bf p_2}|/(m_1+m_2)$ is the relative momentum both in the center-of-mass frame (${\bf p_1+p_2}=0$)~\cite{Song:2016lfv}.
The width parameter $\sigma$ is related to the mean-square radius of the meson through
\begin{eqnarray}
\sigma^2=\frac{4}{3}\frac{(m_1+m_2)^2}{m_1^2+m_2^2}\langle r_D^2\rangle.
\end{eqnarray}

We now consider a static heavy quark located at the origin, 
\begin{eqnarray}
f({\bf r_1},{\bf p_1})=(2\pi)^3\delta^{(3)}({\bf r_1})\delta^{(3)}({\bf p_1}).
\end{eqnarray}

In this limit, Eq.~(\ref{coal}) reduces to the coalescence probability of a static heavy quark,
\begin{eqnarray}
P_3=\frac{1}{(2\pi)^3}\frac{D_3}{2N_c D_2}\int d^3{\bf r_2}d^3{\bf p_2}f_2({\bf r_2,p_2})\Phi(r,k),
\label{coal-c}
\end{eqnarray}
where $D_1=2N_c$.

In the presence of a heavy-quark potential, the distribution function of light antiquark at $T=T_c$ is written as
\begin{eqnarray}
f_2({\bf r_2,p_2})=\frac{D_2}{e^{(\sqrt{m_q^{*2}+p_2^2}+V^*(r_2))/T_c}+1},
\label{f2}
\end{eqnarray}
where $V^*(r)=V(r)-V(\infty)$ is the subtracted potential satisfying $V^*(\infty)=0$. The constant contribution $V(\infty)$ is absorbed into the dressed quark masses~\cite{Gubler:2020hft}:
\begin{eqnarray}
m_Q+m_q+V(\infty)=m_Q^*+m_q^*.
\end{eqnarray}

For example, the charm quark mass increases from $m_c=1.14$ GeV to an effective mass of $m_c^*=1.82$ GeV~\cite{Song:2024rjh}.
Unlike the treatment adopted in Ref.~\cite{Song:2025zfy}, no cutoff is required for the potential term in Eq.~(\ref{f2}), because the Fermi-Dirac distribution suppresses the short-distance contribution, compared to the Boltzmann distribution.

Substituting Eqs. (\ref{wignerF}) and (\ref{f2}) into Eq.~(\ref{coal-c}),
\begin{eqnarray}
P_3&=&\frac{8 D_3}{\pi N_c}\int dr_2 r_2^2\int dp_2 p_2^2\nonumber\\
&\times& \frac{e^{-r^2/\sigma^2-\sigma^2k^2}}{e^{(\sqrt{m_q^{*2}+p_2^2}+V^*(r_2))/T_c}+1},
\label{final}
\end{eqnarray}
where $D_3$ denotes the spin degeneracy of the $D$ meson. The relative momentum in the center-of-mass frame is
\begin{eqnarray}
k^2=\frac{m_Q^{*2}p_2^2}{m_Q^{*2}+m_q^{*2}+2m_Q^*E_2}
\end{eqnarray}
with $E_2=\sqrt{m_q^{*2}+p_2^2}$.

Ignoring the potential $V^*(r)$, the spatial part can be integrated out:
\begin{eqnarray}
\int d^3{\bf r_2}\exp\bigg[-\frac{r^2}{\sigma^2}\bigg]=\frac{1}{\gamma_{c.m.}}\int d^3{\bf r}\exp\bigg[-\frac{r^2}{\sigma^2}\bigg]\nonumber\\
=\frac{\sqrt{m_Q^{*2}+m_q^{*2}+2m_Q^*E_2}}{m_Q^*+E_2}(\sqrt{\pi}\sigma)^3,
\label{limit2}
\end{eqnarray}
where $\gamma_{c.m.}$ is the gamma factor of center-of-mass in heat bath frame, and the coalescence probability is simplified into~\cite{Song:2021mvc} 
\begin{eqnarray}
P_3
=\frac{2\sigma^3D_3}{N_c\sqrt{\pi}} \int dp_2p_2^2\frac{e^{-\sigma^2k^2}}{\gamma_{c.m.}(e^{\sqrt{m_2^2+p_2^2}/T}+1)}.
\end{eqnarray}

For pseudoscalar $D$ meson, the spin degeneracy factor ($D_3$ in Eq.~(\ref{final}) equals 1, whereas for $D^*$ meson it is 3.
In principle, the coalescence of all charm hadrons can be evaluated using the corresponding Wigner functions and hadronic radii~\cite{Cho:2014xha,Kordell:2021prk,Zhao:2023dvk}.
In practice, however, one may assume that the statistical hadronization model accurately describes the relative abundances of all charm hadrons~\cite{Andronic:2021erx} and focus exclusively on the coalescence of the ground-state pseudoscalar $D$ meson~\cite{Song:2021mvc}.
The total charm-quark coalescence probability can then be estimated by multiplying the $D$-meson coalescence probability by the ratio of the total charm-hadron density to the $D^0$ density at $T_c$~\cite{Song:2021mvc},
\begin{eqnarray}
\frac{n^{\rm charm~hadron}(T_c)}{n^{D^0}(T_c)}=7.5,
\label{ratio-charm}
\end{eqnarray}
where $n^i(T)$ denotes the equilibrium number density of species $i$ at temperature $T$ in grand canonical ensemble.

\begin{figure}[h]
\centerline{
\includegraphics[width=9 cm]{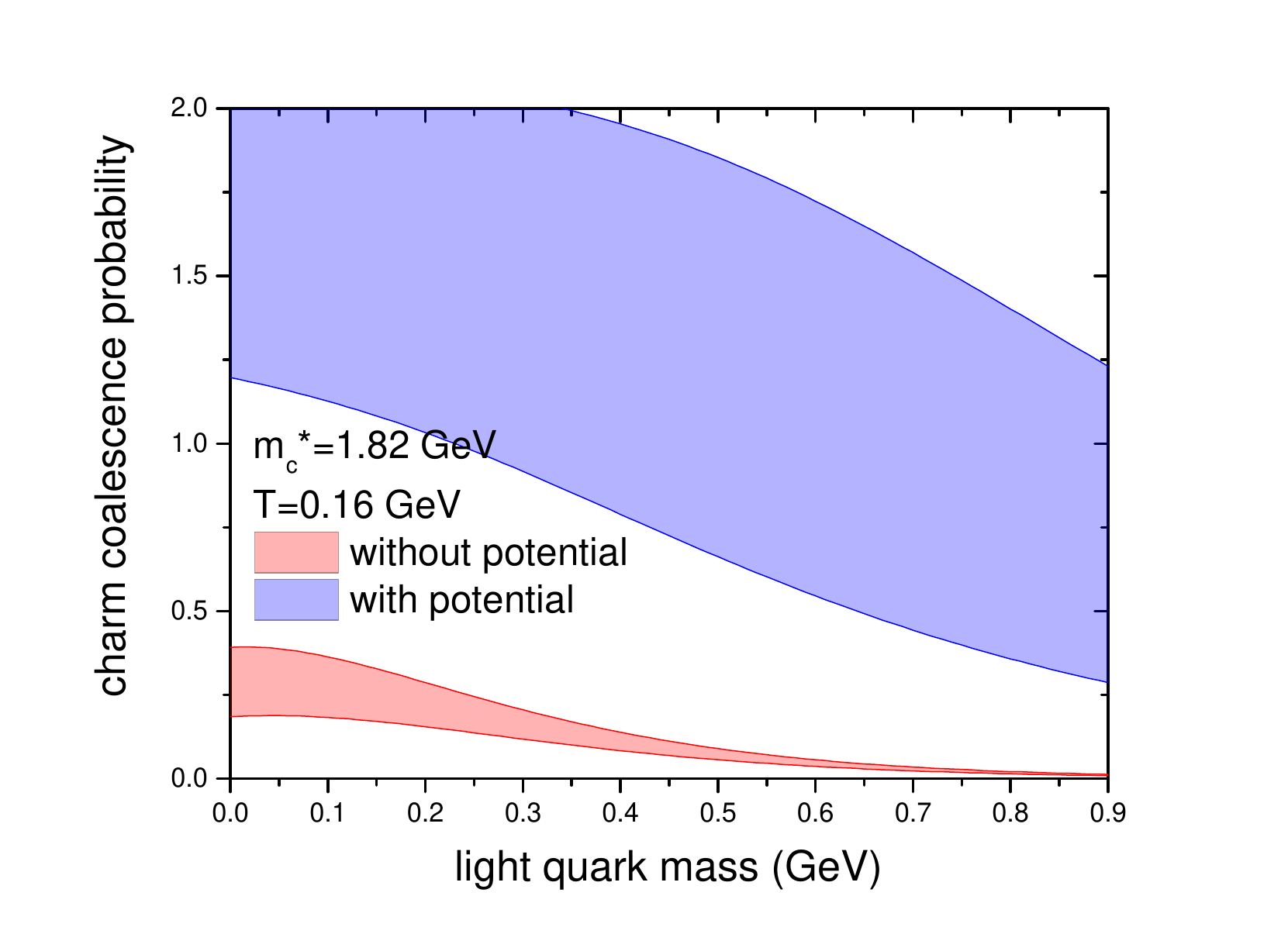}}
\caption{Charm-quark coalescence probabilities at $T=$ 0.16 GeV (blue) with and (red) without potential as a function of the light quark mass, assuming $D$ meson radius is between 0.5 fm and 1.0 fm.}
\label{V0}
\end{figure}

Figure~\ref{V0} presents the total charm-quark coalescence probability, obtained from Eq.~(\ref{final}) multiplied by the factor in Eq.~(\ref{ratio-charm}), as a function of the light (anti)quark mass.
Since the $D$-meson radius is not precisely known, we consider values between 0.5 and 1.0 fm.
 
Without the heavy-quark potential, the total coalescence probability remains substantially below unity.
This result suggests that charm-quark hadronization cannot be described solely by recombination with thermal light quarks and anti-quarks that are uniformly distributed in coordinate space.
In contrast, including the heavy-quark potential significantly enhances the coalescence probability, allowing it to approach unity over a broad range of parameters.

Interestingly, after the potential is included, the ordering of the uncertainty band is reversed: the upper boundary corresponds to a $D$-meson radius of 0.5 fm, whereas the lower boundary corresponds to 1.0 fm. This behavior arises because the attractive heavy-light quark potential concentrates light anti-quarks near the heavy quark in coordinate space, thereby enhancing the coalescence probability more strongly for smaller meson radii. 

According to lattice QCD calculations~\cite{Borsanyi:2010cj,Borsanyi:2012cr}, the energy density at the crossover temperature with vanishing baryon chemical potential is approximately ${\rm 0.4~ GeV/fm^3}$.
Within the quasiparticle picture, this energy density corresponds to an effective light-quark mass of about 0.45 GeV~\cite{Plumari:2011mk,Moreau:2019vhw}.
Taking into account the mass difference between $D$ and $D_s$ mesons, the corresponding strange-quark mass is approximately 0.55 GeV.
We neglect explicit gluonic degrees of freedom by assuming that the gluons have already converted into quark and anti-quark pairs near $T_c$. 

\begin{figure}[h]
\centerline{
\includegraphics[width=9 cm]{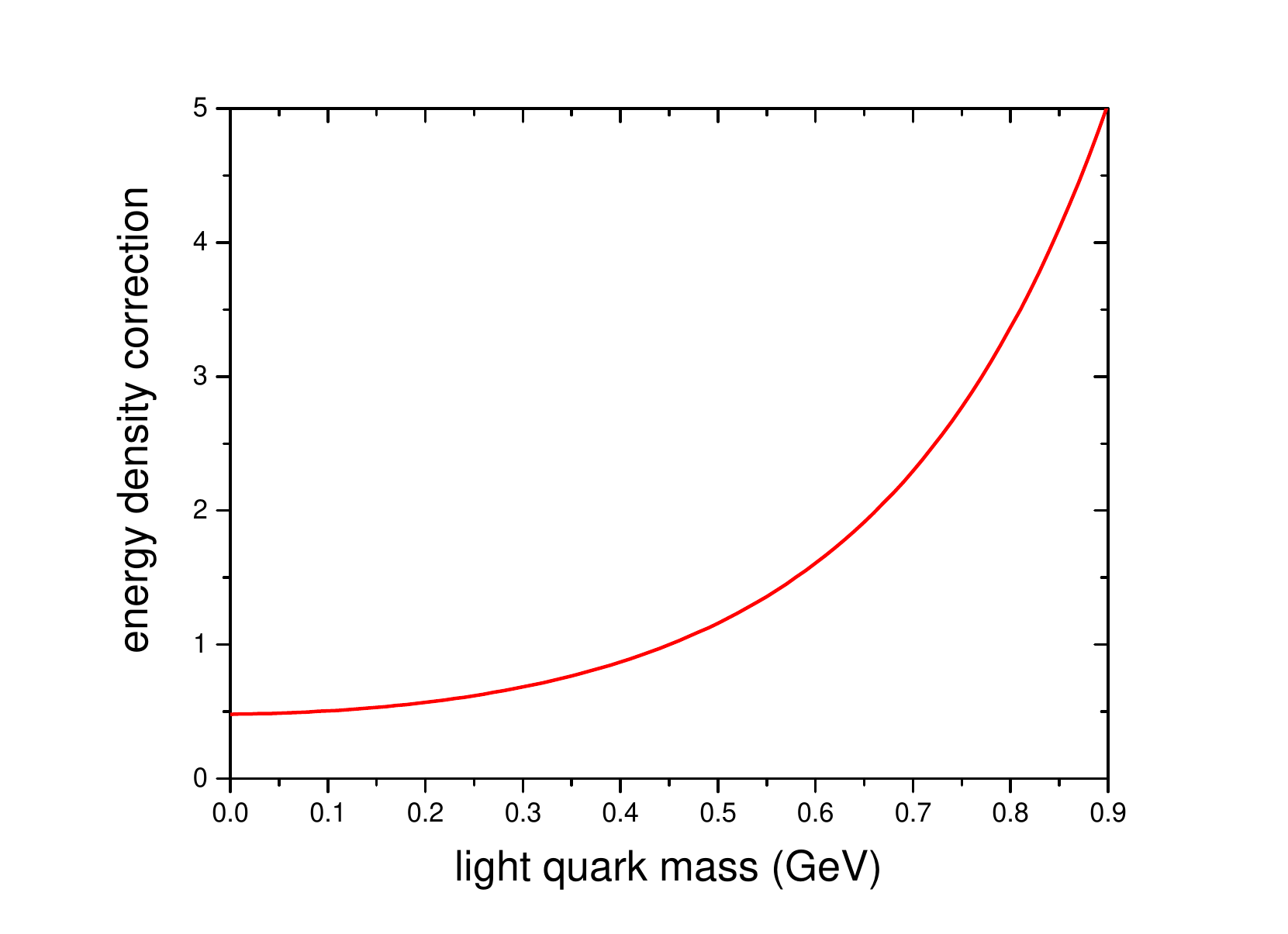}}
\centerline{
\includegraphics[width=9 cm]{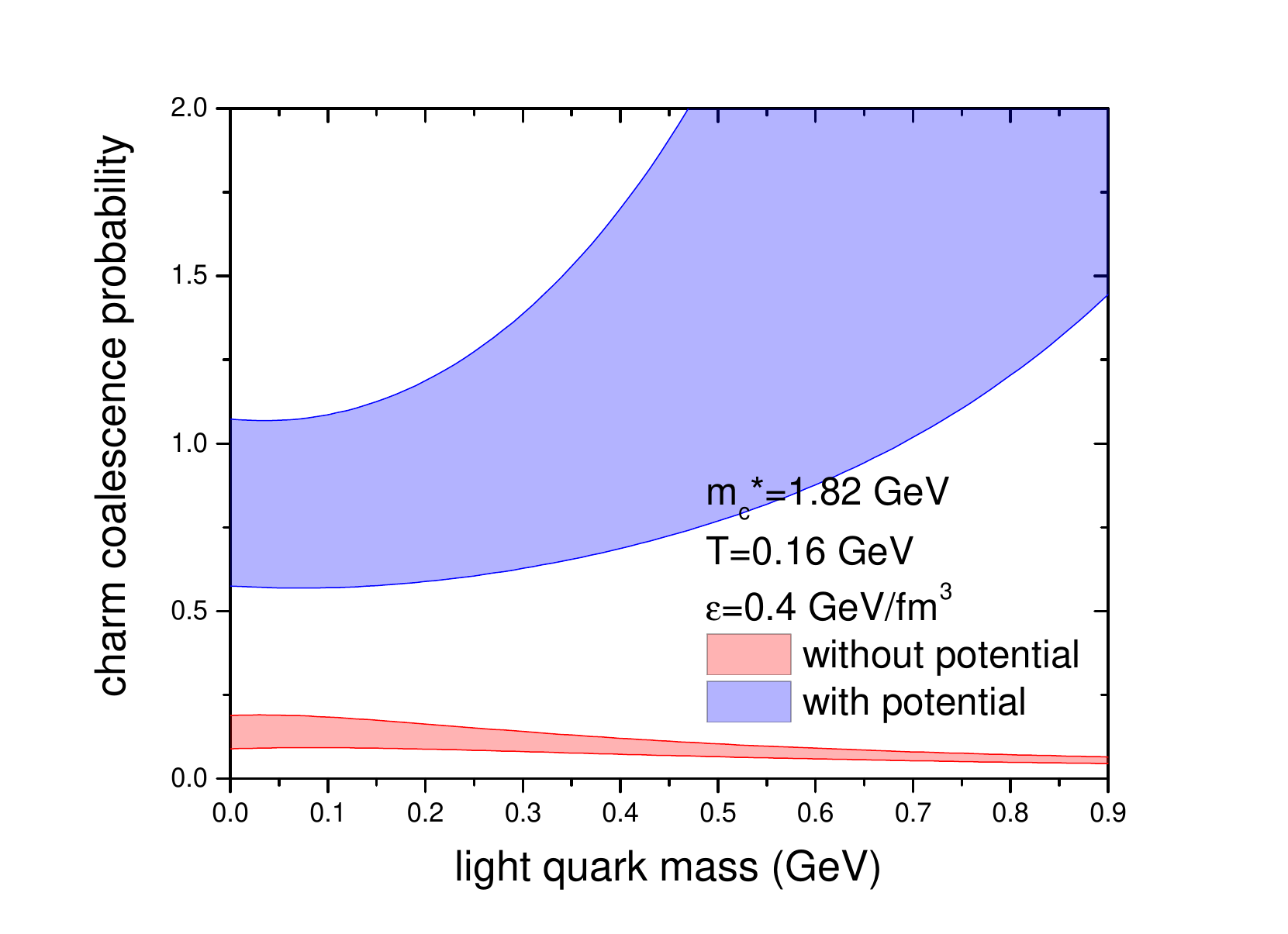}}
\caption{(Upper) the energy density correction and (lower) charm quark coalescence probabilities at $T=$ 0.16 GeV with and without potential as a function of the light quark mass, considering the energy density from lattice QCD calculations~\cite{Borsanyi:2010cj,Borsanyi:2012cr}.}
\label{coal-pot2}
\end{figure}

To maintain consistency with the lattice QCD equation of state, we introduce the correction factor shown in the upper panel of Fig.~
\ref{coal-pot2}, such that the total energy density remains fixed at 0.4 ${\rm GeV/fm^3}$ for all values of the light-quark mass.
The resulting coalescence probabilities are displayed in the lower panel of Fig.~\ref{coal-pot2}.
After applying this correction, the coalescence probability is suppressed at small light-quark masses and enhanced at larger masses.
Nevertheless, the probability obtained without the heavy-quark potential remains well lower than unity, consistent with the findings of Ref.~\cite{Song:2021mvc}, whereas the inclusion of the potential yields a coalescence probability close to unity.

The same analysis can be performed for bottom quarks.
The effective bottom-quark mass is taken to be $m_b^*=5.17$ GeV~\cite{Song:2024rjh}.
The ratio of the total bottom-hadron density to that of the ground-state $B^-$ meson is taken to be~\cite{Song:2021mvc}:
\begin{eqnarray}
\frac{n^{\rm bottom~hadron}(T_c)}{n^{B^-}(T_c)}=11.4.
\label{ratio-bottom}
\end{eqnarray}

\begin{figure}[h]
\centerline{
\includegraphics[width=9 cm]{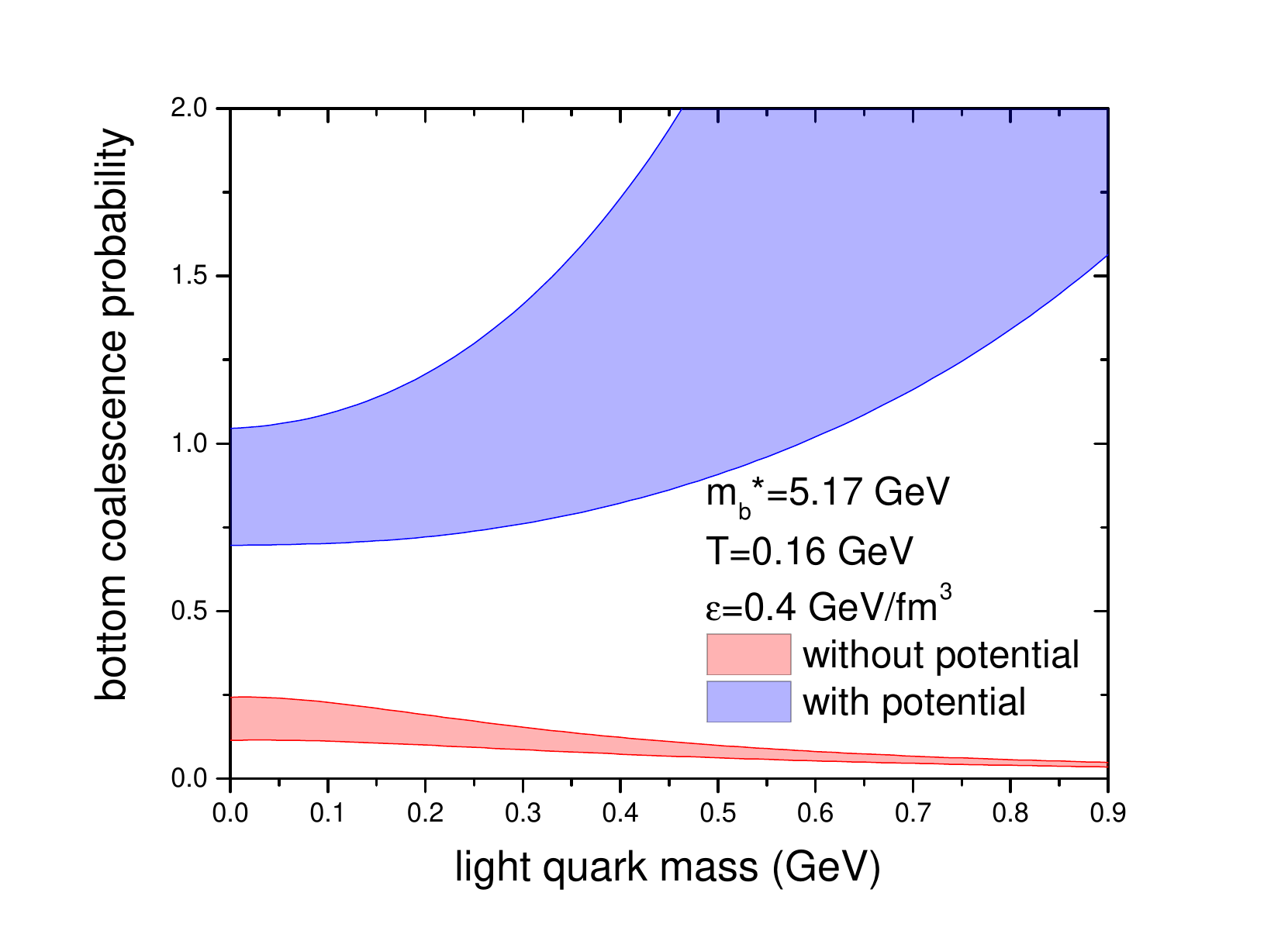}}
\caption{Bottom-quark coalescence probabilities at $T=$ 0.16 GeV with and without potential as a function of the light quark mass. $B$ meson radius varies from 0.5 fm to 1.0 fm.}
\label{coal-pot2-b}
\end{figure}

Figure~\ref{coal-pot2-b} shows the bottom-quark coalescence probability at $T_c$ as a function of the light-anti-quark mass for $B$-meson radius in the range 0.5-1.0 fm, after imposing the lattice-QCD energy-density constraint. 
Similar to the charm sector, the coalescence probability obtained without the heavy-quark potential is substantially underestimated.
Once the potential is included, however, the coalescence probability is significantly enhanced and approaches unity.

\section{coalescence with in-medium potential}
\label{in-medium}
So far, the heavy-quark coalescence probabilities have been calculated using the vacuum heavy-light quark potential, corresponding to $\mu/\sqrt{\sigma}\approx 0.35$.

In the QGP, however, the interaction potential is expected to be modified by color screening.
Within the present framework, this medium effect is incorporated through the screening mass $\mu$ appearing in Eq.~(\ref{FcFs}).

\begin{figure}[h]
\centerline{
\includegraphics[width=9 cm]{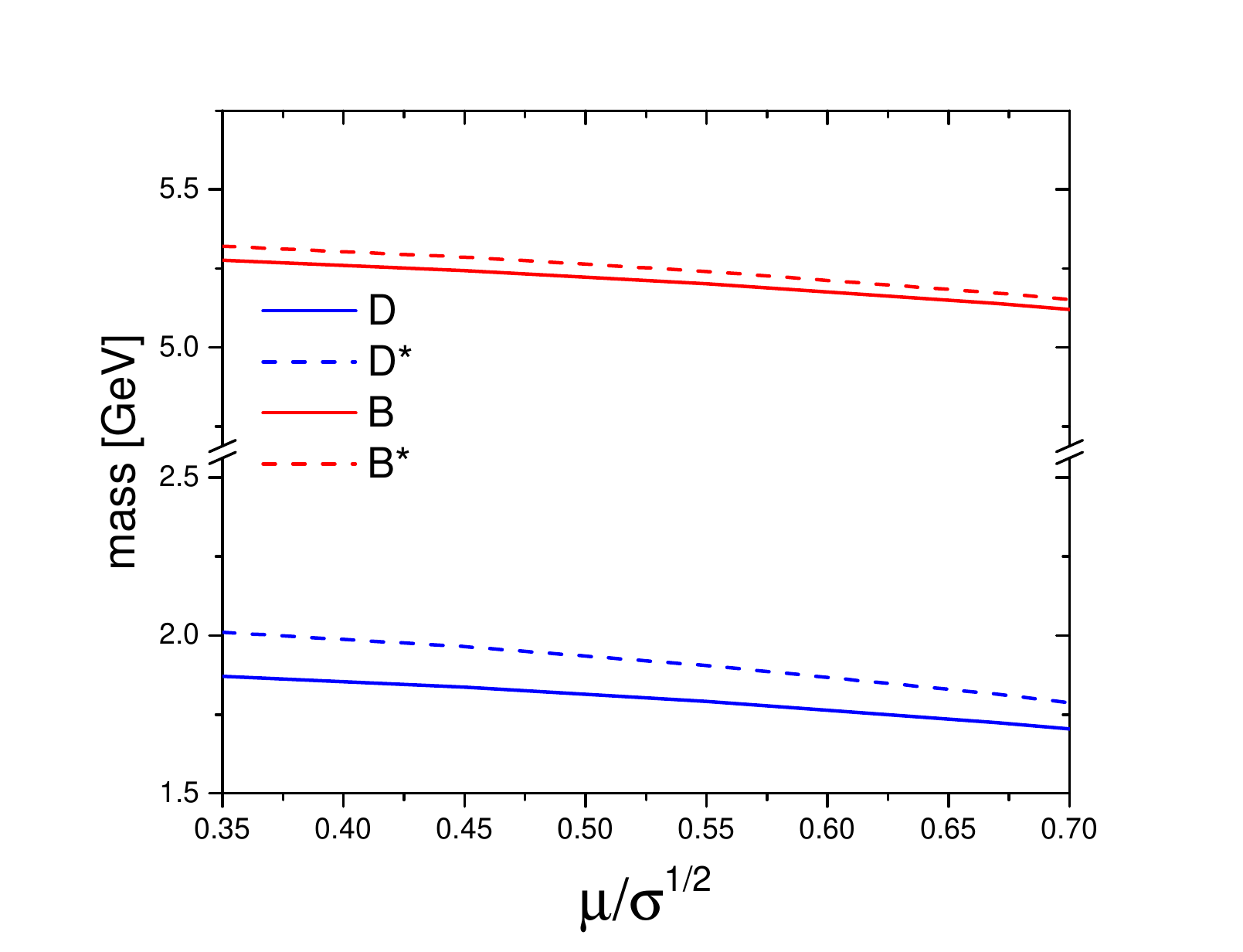}}
\caption{Masses of $D$, $D^*$, $B$ and $B^*$ mesons in the QGP as functions of the screening parameter $\mu/\sqrt{\sigma}$}
\label{muB}
\end{figure}

Figure~\ref{muB} shows the masses of the $D$, $D^*$, $B$ and $B^*$ mesons as functions of the screening mass $\mu$.
As expected, the meson masses decrease with increasing $\mu$.
For example, when $\mu/\sqrt{\sigma}$ increases from 0.35 to 0.5, the $D$-meson mass decreases from 1.87 GeV to 1.81 GeV, while the $B$-meson mass decreases from 5.28 GeV to 5.22 GeV. 
The resulting mass shifts are relatively modest, amounting to approximately 60 MeV for both systems.

\begin{figure}[h]
\centerline{
\includegraphics[width=9 cm]{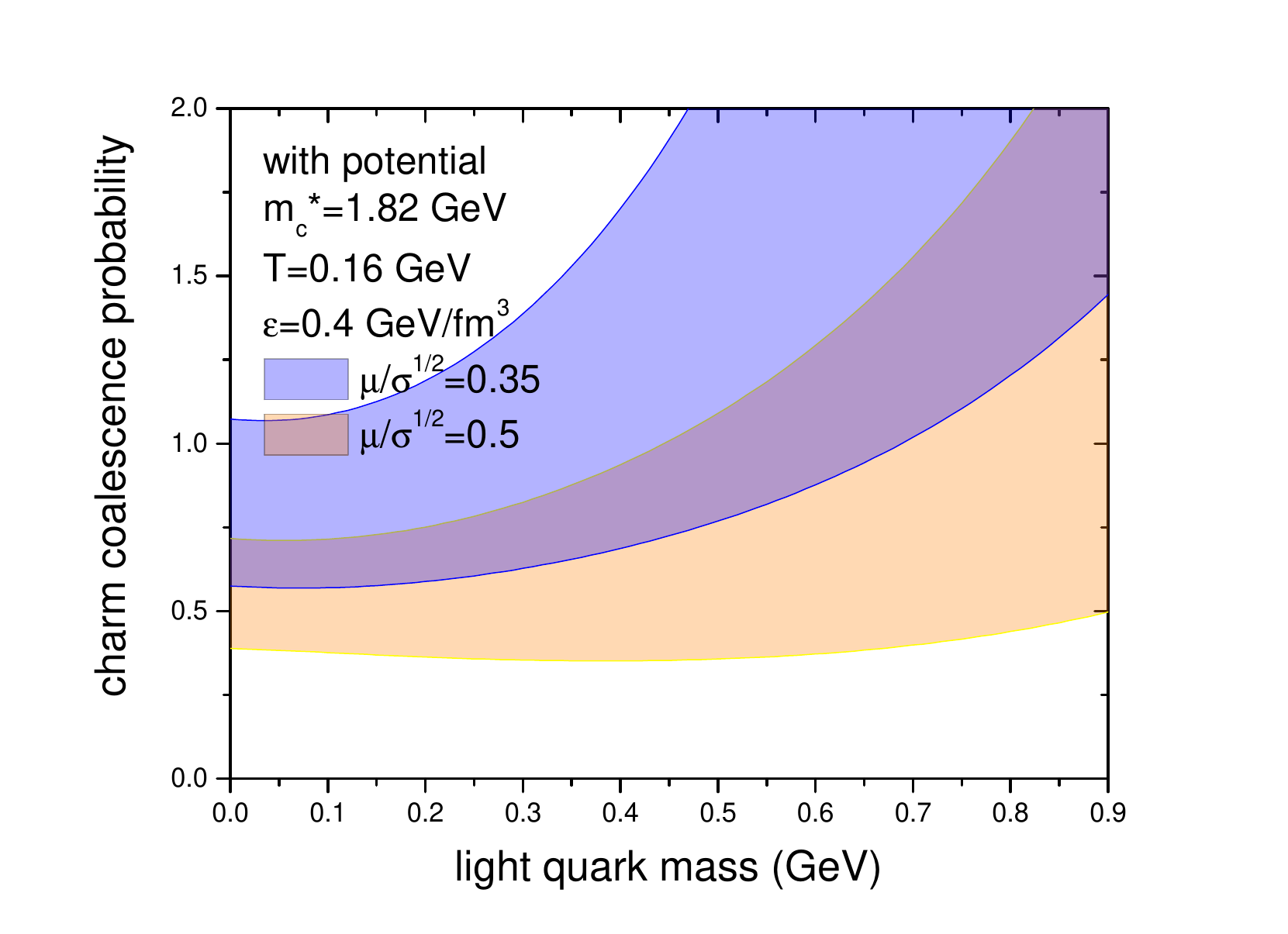}}
\centerline{
\includegraphics[width=9 cm]{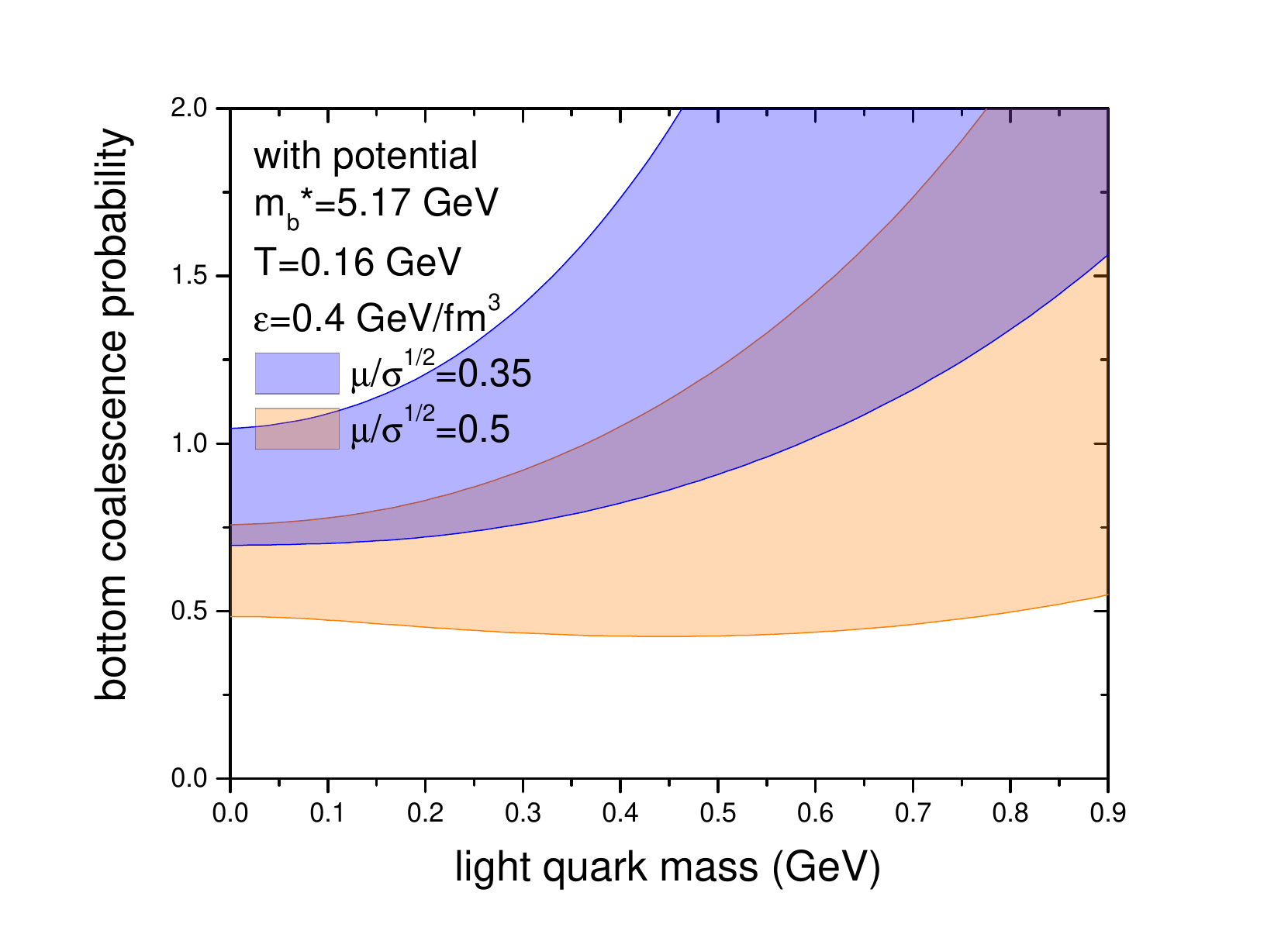}}
\caption{Charm-quark (upper panel) and bottom-quark (lower panel) coalescence probabilities at $T_c$ as functions of the light-quark mass for $\mu/\sqrt{\sigma}=$ 0.35 and 0.5, including the lattice-QCD energy -density constraint. The upper and lower boundaries of the shaded bands correspond to heavy-meson radii of 0.5 fm and 1.0 fm, respectively.}
\label{mu}
\end{figure}

Although the masses of $D$ and $B$ mesons exhibit only a weak dependence on the screening mass, the coalescence probabilities are considerably more sensitive to this parameter.
Figure~\ref{mu} presents the charm- and bottom-quark coalescence probabilities at $T_c$ for $\mu/\sqrt{\sigma}=$ 0.35 and 0.5 as functions of the light-quark mass.
The blue bands correspond to the results shown previously in Figs.~\ref{coal-pot2} and \ref{coal-pot2-b}, while the orange bands represent the results obtained with $\mu/\sqrt{\sigma}=$ 0.5.

The coalescence probability decreases as the screening mass increases.
This behavior can be understood from the weakening of the heavy-light quark interaction: stronger color screening reduces the depth of the potential well and consequently lowers the spatial concentration of light anti-quarks around the heavy quark.
As in Figs.~\ref{coal-pot2} and \ref{coal-pot2-b}, the upper and lower boundaries of the shaded bands correspond to heavy-meson radii of 0.5 fm and 1.0 fm, respectively.

For the light-quark mass favored by the quasiparticle description of the lattice-QCD equation of state, $m_q \simeq 0.45$ GeV, the charm- and bottom-quark coalescence probabilities remain close to unity for a heavy-meson radius of 0.5 fm when $\mu/\sqrt{\sigma}=0.5$.
For larger screening masses, however, the coalescence probability falls below unity even for this smallest radius considered.
This observation suggests that the screening mass near $T_c$ is unlikely to exceed $\mu/\sqrt{\sigma}\approx 0.5$.

At this value of the screening parameter, the masses of the $D$ and $B$ mesons are reduced by approximately 60 MeV relative to their vacuum values.
This estimate is consistent with predictions from effective Lagrangian approaches and QCD sum-rule calculations~\cite{Montana:2020lfi,Gubler:2020hft}.

\section{summary}
\label{summary}
The parton coalescence model is one of the most successful frameworks for describing parton hadronization near the QGP phase-transition boundary.
Its validity is supported by the observed constituent-quark-number scaling of elliptic flow in relativistic heavy-ion collisions.
However, when the light anti-quarks that serve as coalescence partners of heavy quarks are assumed to be uniformly distributed in coordinate space, the resulting total heavy-quark coalescence probability at $T_c$ is significantly below unity.

In this work, the heavy-light quark potential that reproduces the masses of both pseudoscalar and vector heavy mesons.
The attractive interaction enhances the local density of light anti-quarks around a heavy quark and thereby increases the coalescence probability.
As a result, the total coalescence probability can approach unity without introducing an ad hoc normalization factor.

In the QGP, the heavy-light quark potential is expected to be modified by color screening.
To investigate this effect, we varied the screening mass appearing in the potential.
We found that the masses of heavy mesons decrease with increasing screening mass.
At the same time, resulting in a shallower potential and a corresponding reduction in the heavy-quark coalescence probability.

Requiring the total coalescence probability to remain close to unity for a static heavy quark places a constraint on the in-medium modification of the potential.
Within the present framework, values of the screening parameter larger than $\mu/\sqrt{\sigma}\simeq 0.5$ lead to a substantial suppression of the coalescence probability. 
This value corresponds to a reduction of approximately 60 MeV in the masses of the $D$ and $B$ mesons relative to their vacuum values. 
Our results therefore suggest that the in-medium mass shifts of open heavy-flavor mesons near $T_c$ are unlikely to exceed this magnitude.

\section*{Acknowledgements}
We are grateful for useful discussions with E. Bratkovskaya.
We acknowledge support by the Deutsche Forschungsgemeinschaft (DFG, German Research Foundation) through the grant CRC-TR 211 'Strong-interaction matter under extreme conditions' - Project number 315477589 - TRR 211. 
The computational resources have been provided by the LOEWE-Center for Scientific Computing and the "Green Cube" at GSI, Darmstadt and by the Center for Scientific Computing (CSC) of the Goethe University.

\bibliography{main}

\end{document}